\documentclass[apjl]{emulateapj}

\slugcomment{}

\shorttitle{A New Estimate for Black Hole Mass}
\shortauthors{Minezaki et al.}

\begin{document}

\title{A NEW BLACK HOLE MASS ESTIMATE FOR OBSCURED ACTIVE GALACTIC NUCLEI}

\author
{Takeo Minezaki\altaffilmark{1}
and Kyoko Matsushita\altaffilmark{2}
}

\altaffiltext{1}{Institute of Astronomy, School of Science, University
of Tokyo, 2-21-1 Osawa, Mitaka, Tokyo 181-0015, Japan; minezaki@ioa.s.u-tokyo.ac.jp}
\altaffiltext{2}{Department of Physics, Tokyo University of Science,
1-3 Kagurazaka, Shinjuku-ku, Tokyo 162-8601, Japan}

\begin{abstract}
We propose a new method for estimating the mass of
a supermassive black hole,
applicable to obscured Active Galactic Nuclei (AGNs).
This method estimates the black hole mass using
the width of the narrow core of the neutral FeK$\alpha $ emission line
in X-rays and the distance of its emitting region from the black hole
based on the isotropic luminosity indicator
via the luminosity scaling relation.
We collect the line width data of the neutral FeK$\alpha $ line core
for seven type-1 AGNs and seven type-2 AGNs
obtained by the Chandra High Energy Transmission Grating Spectrometer,
which affords the best spectral resolution currently available.
Assuming the virial relation between the locations and the velocity widths of
the neutral FeK$\alpha $ line core and the broad H$\beta $ emission line,
the luminosity scaling relation of 
the neutral FeK$\alpha $ line core emitting region
is estimated.
We find that the full width at half maximum
of the neutral FeK$\alpha $ line core falls between 
that of the broad Balmer emission lines and
the corresponding value at the dust reverberation radius
for most of the type-1 AGNs and for all of the type-2 AGNs.
This suggests that significant fraction of photons of
the neutral FeK$\alpha $ line core originates between
the outer BLR and the inner dust torus in most cases.
The black hole mass $M_{\rm BH,FeK\alpha}$ estimated with this method
is then compared with other black hole mass estimates,
such as the broad emission-line reverberation mass $M_{\rm BH,rev}$
for the type-1 AGNs,
the mass $M_{\rm BH,H_2O}$ based on the H$_2$O maser
and
the single-epoch mass estimate $M_{\rm BH,pol}$
based on the polarized broad Balmer lines
for the type-2 AGNs.
We find that $M_{\rm BH,FeK\alpha}$ is consistent with
$M_{\rm BH,rev}$ for the most of the type-1 AGNs
and with $M_{\rm BH,pol}$ for all of the type-2 AGNs.
We also find that $M_{\rm BH,FeK\alpha}$
is correlated well with $M_{\rm BH,H_2O}$ for the type-2 AGNs.
These results suggest that $M_{\rm BH,FeK\alpha}$ is a potential
indicator of the black hole mass especially for obscured AGNs.
In contrast, $M_{\rm BH,FeK\alpha}$
for which the same virial factor as for $M_{\rm BH,rev}$ and $M_{\rm BH,pol}$ is adopted
is systematically larger than $M_{\rm BH,H_2O}$ by about a factor of $\sim 5$,
and the possible origins are discussed.
\end{abstract}


{\small
$Online$-$only\ material:$ color figures
}

\section{INTRODUCTION}

The correlation between the mass of the supermassive
black holes in the centers of galaxies
and the bulge stellar mass or the velocity dispersion
\citep[e.g.,][]{1995ARA&A..33..581K,1998AJ....115.2285M,2013ARA&A..51..511K}
strongly suggest that
the growth of the central supermassive black hole and
the evolution of the galaxies are connected.
Obscured Active Galactic Nuclei (AGNs) are important targets
for studying the AGN-galaxy coevolution because
a large number of host galaxies
can be investigated in detail being free from
the strong central emission from the AGN itself.
Moreover, they are of special importance
because some types of obscured AGNs
such as the ultraluminous infrared galaxies (ULIRGs)
have long been anticipated to be
in the middle of the evolutionary sequence of AGNs
\citep[e.g.,][]{1988ApJ...325...74S,1996ARA&A..34..749S,2008ApJS..175..356H}.
The ``new type AGNs'' recently found by \citet{2007ApJ...664L..79U},
where the central engine is thought to be deeply obscured
by a dust torus with an unusually small opening angle,
are also interesting for its nature and role 
in the AGN evolution.

Consequently, measuring the black hole mass in obscured AGNs is important;
however, the precise measurement is not straightforward.
The position-velocity diagram of the molecular disk
surrounding the black hole,
obtained by the VLBI observation of the H$_2$O maser,
is considered to be one of the most
precise estimate for the black hole mass \citep[e.g.,][]{1995Natur.373..127M}.
However, precise alignment between the equatorial plane of
the molecular disk and the observer's line of sight is required
for the disk H$_2$O maser to be observable;
thus the number of applicable targets is very limited.
The single-epoch mass estimate based on the luminosity scaling relation
of the broad emission-line region (BLR)
\citep[e.g.,][]{1998ApJ...505L..83L,2002MNRAS.337..109M,2006ApJ...641..689V}
measuring the widths of the polarized broad Balmer emission lines
would be widely available for type-2 AGNs.
However, a significant fraction of type-2 AGNs does not
present the polarized broad Balmer emission lines
\citep[][]{2001ApJ...554L..19T}.
Moreover, this type of measurement might be unavailable
for the ULIRGs and the new-type AGNs,
whose BLR is largely covered by obscuring materials.
A precise estimate of the black hole mass 
that is widely available for obscured AGNs is desired.

The fluorescent iron K$\alpha $ line 
(hereafter, FeK$\alpha $ line)
is an almost ubiquitous emission line observable
for both obscured and unobscured AGNs
\citep[][]{2004ApJ...604...63Y,2007MNRAS.382..194N,2011ApJ...727...19F}.
The rest-frame energy of the narrow core of the FeK$\alpha $ emission line
is close to that of neutral or low-ionization-state irons. 
Considering its narrow line width,
the emitting region is supposed to be not very close to the black hole,
such as the outer parts of the accretion disk and more distant matter.
The large equivalent width indicates both large covering fraction
and large optical depth for the emitting region,
which suggests that 
the neutral FeK$\alpha $ line core originates in the dust torus
surrounding the central engine
in accordance with the unified model of AGNs \citep[e.g.,][]{1991PASJ...43..195A}.

When the width of an emission line and
the distance of its emitting region from the black hole are obtained,
the black hole mass can be estimated assuming the virial relation.
\citet{2011SCPMA..54.1354J} assumed that
the location of the emitting region of the neutral FeK$\alpha $ line core
corresponds to the near-infrared $K$-band reverberation radius
for the inner region of the dust torus,
and estimated the black hole mass of 10 type-1 AGNs
on the basis of the FeK$\alpha $ line widths
and the luminosity scaling relation of the dust reverberation radius
to the optical $V$-band luminosity by \citet{2006ApJ...639...46S}.
They examined the black hole mass on the basis of 
the neutral FeK$\alpha $ line core
by comparing the black hole mass obtained from
the broad emission-line reverberation;
however, the correlation between them was statistically insignificant,
which could be attributable to small sample size and large uncertainties
in the FeK$\alpha $ line widths.

Recently, \citet{gree+10} and \citet{kosh+14} presented
the luminosity scaling relations
of the broad H$\beta $ emission-line reverberation radius
and the dust reverberation radius
to isotropic luminosity indicators
that are unbiased by the obscuration,
such as
the luminosity of the [\ion{O}{4}] $\lambda 25.89\ \mu$m emission line
$L_{{\rm [OIV]}}$,
and that of the {\it Swift} BAT hard X-ray (14--195 keV) band $L_{\rm BAT}$
\citep[e.g.,][]{mele+08,2009ApJ...700.1878R,diam+09}.
Using these radius-luminosity scaling relations,
it becomes possible to derive the radius of
the emitting region of the neutral FeK$\alpha $ line core
for obscured AGNs.

In this paper, 
we estimate the black hole mass of both obscured and unobscured AGNs
on the basis of the line width of the neutral FeK$\alpha $ line core
and the luminosity scaling relation of its emitting region radius
with the [\ion{O}{4}] $\lambda 25.89\ \mu$m emission-line luminosity.
Then, the estimated black hole mass was examined by comparing them with
other black hole mass estimates that have been established.
We assume the cosmology of
$H_0=73$ km s$^{-1}$ Mpc$^{-1}$,
$\Omega _m = 0.27$, and $\Omega _\Lambda = 0.73$
according to \citet{2007ApJS..170..377S} throughout this paper.

\section{TARGETS}
We collect the line width data of
the neutral FeK$\alpha $ line core obtained by
the {\it Chandra} High Energy Transmission Grating Spectrometer
\citep[HETGS;][]{1994SPIE.2280..168M}, which affords
the best spectral resolution currently available in the FeK$\alpha$ 
energy band ($\sim 39$ eV, or $\sim 1860$ km s$^{-1}$
in full width at half maximum; FWHM),
and we further select target AGNs with other
mass estimates of their central supermassive black holes.

For the type-1 AGNs, seven target AGNs
whose black hole masses are measured by the reverberation
of the broad emission lines
are selected out of 12 AGNs,
for which the best FWHM constraint for 
the neutral FeK$\alpha $ line core 
was obtained by \citet{2010ApJS..187..581S}.
The type-1 target AGNs and
their data are listed in Table \ref{data_sy1}.

For the type-2 AGNs, four target AGNs
whose black hole masses are measured by
the VLBI observation of the H$_2$O maser
are selected out of eight AGNs,
for which the FWHM of the neutral FeK$\alpha $ line core
was measured by \citet{2011ApJ...738..147S}.
In addition, one type-2 AGN from the 12 AGNs of \citet{2010ApJS..187..581S}
and five type-2 AGNs from the eight AGNs of \citet{2011ApJ...738..147S} 
are also selected,
for which the FWHMs of the polarized broad Balmer
emission lines are available to obtain
the single-epoch black hole mass estimate
on the basis of the luminosity scaling relation of the BLR.
In total, seven type-2 AGNs are selected
because three objects in the $4+6$ type-2 AGNs are common.
The type-2 target AGNs and their data are listed in Table \ref{data_sy2}.

\section{RESULTS}
\subsection{Velocity Width of the Neutral FeK$\alpha $ Line Core}
First, we compare the velocity width of the neutral FeK$\alpha $ line core
with that of the broad Balmer emission lines,
and that corresponding to the dust reverberation radius
to examine the location of its emitting region.

In Figure \ref{fwhm_fwhm}, we plotted
the FWHM of the neutral FeK$\alpha $ line core
against those of the broad H$\beta $ emission line
and the polarized broad Balmer emission lines
for the type-1 and type-2 target AGNs, respectively.
We also plotted the line that represents
the FWHM at the dust reverberation radius observed in the $K$ band,
which is estimated by scaling the FWHM of
the broad H$\beta $ emission line according to the virial relation
based on the systematic difference of 
the reverberation radii of the broad H$\beta $ emission line
and the near-infrared dust emission \citep[][]{kosh+14}.

As shown in Figure \ref{fwhm_fwhm},
we find that the FWHM of the neutral FeK$\alpha $ line core
falls between that of the broad Balmer emission lines
and the corresponding value at the dust reverberation radius
for all of the type-2 AGNs, and for most of the type-1 AGNs
except the only one outlier of NGC 7469.
This suggests that
significant fraction of photons of the neutral FeK$\alpha $ line core
originate between the outer BLR and the inner dust torus in most cases.

\subsection{Luminosity Scaling Relation of the Neutral FeK$\alpha $ Line Core Emitting Region}

As a next step,
we scaled the radius-luminosity scaling relation of
the broad H$\beta $ emission line
according to the systematic difference between
the FWHMs of the neutral FeK$\alpha $ line core
and the broad H$\beta $ emission line for the type-1 AGNs
assuming the virial relation.
It is estimated
by the linear regression analysis where the slope is fixed to unity,
and
the best-fit linear regression is
$\log {\rm FWHM_{FeK\alpha }}=\log {\rm FWHM_{H\beta}}-0.186\ (\pm 0.088)$
with an additional scatter of 0.11 dex in both directions
for the reduced $\chi ^2$ to achieve unity.
Then, the emitting region radius of the neutral FeK$\alpha $ line core
is estimated as
\begin{equation}
\log {\rm r_{FeK\alpha }}=\log {\rm r_{H\beta}}-0.37\ (\pm 0.18)
\label{r_fekahb}
\end{equation}
according to the virial relation.
\citet{kosh+14} estimated the luminosity scaling relation of
the reverberation radii to the luminosity of
the [\ion{O}{4}] emission line in the form of 
$\log r = \alpha + 0.5 \log L_{{\rm [OIV]}}/10^{41}\ {\rm erg}\ {\rm s}^{-1}$,
as
$\alpha = -1.94\ (\pm 0.07)$ and $\alpha = -1.28\ (\pm 0.07)$
for 
the reverberation radius of the broad H$\beta $ emission line
and
the dust reverberation radius observed in the $K$ band,
respectively.
By scaling the former relation
according to Equation (\ref{r_fekahb}),
the luminosity scaling relation of the emitting region radius
of the neutral FeK$\alpha $ line core is estimated as
\begin{equation}
\log r_{{\rm FeK\alpha}} = -1.57\ (\pm 0.19) + 0.5\log L_{{\rm [OIV]}}/10^{41}\ {\rm erg}\ {\rm s}^{-1}\ .
\label{r_loiv}
\end{equation}

The radius-luminosity scaling relation
for the neutral FeK$\alpha $ line core 
of Equation (\ref{r_loiv})
is located between the reverberation radii
of the broad H$\beta $ emission line
and the dust torus emission in the $K$ band
as indicated by Figure \ref{fwhm_fwhm},
and we use it
for the estimation of the black hole mass
in the next section.

\subsection{Estimating the Black Hole Mass from the Neutral FeK$\alpha $ Line Core}

Next, 
 we estimate the black hole mass
of both type-1 and type-2 target AGNs
using the neutral FeK$\alpha $ line core.
The emitting region radius $r_{\rm FeK\alpha}$ is estimated
from $L_{{\rm [OIV]}}$ and the radius-luminosity scaling relation of Equation (\ref{r_loiv})
because the hard X-ray flux seems to be attenuated
in some heavily obscured AGNs in the targets.
The $L_{{\rm [OIV]}}$ value is taken from \citet{liuw+10},
on the basis of the original data of \citet{mele+08} and \citet{diam+09}
for most of the targets;
however, it is recalculated from the [\ion{O}{4}] emission-line flux of \citet{diam+09},
assuming the redshift-independent distance 
taken from the NASA/IPAC Extragalactic Database for very nearby objects.
In addition, the $L_{{\rm [OIV]}}$ of NGC 2110 
is calculated from the data of \citet{2010ApJ...716.1151W}.
We then estimate the black hole mass assuming the virial relation as
\begin{equation}
M_{\rm BH,FeK\alpha} = f\frac{r_{\rm FeK\alpha}\sigma _{\rm FeK\alpha }^2}{G}
\label{eqn_virial}
\end{equation}
where $f$ is the virial factor and
$\sigma_{\rm FeK\alpha}$ is the velocity dispersion of the neutral FeK$\alpha $ line core.
We adopt $f=5.5$ following the references of the broad emission-line reverberation
for the type-1 target AGNs
\citep[][]{2004ApJ...615..645O,2004ApJ...613..682P,bent+06,bent+07,denn+10,grie+12}.
The velocity dispersion is calculated as $\sigma _{\rm FeK\alpha}={\rm FWHM_{\rm FeK\alpha}}/2.35$
because a Gaussian emission-line profile is used for the spectral fitting
of the neutral FeK$\alpha $ line core in
\citet{2010ApJS..187..581S,2011ApJ...738..147S}.
The $L_{{\rm [OIV]}}$ and the $M_{\rm BH,FeK\alpha}$ for the type-1 and type-2 target AGNs
are listed in Tables \ref{data_sy1} and \ref{data_sy2}, respectively.

The FWHM of the neutral FeK$\alpha$ line core
spreads between that of the broad H$\beta $ line
and that corresponding to the dust reverberation radii in most cases.
This scatter causes a systematic uncertainty in the $r_{\rm FeK\alpha}$
derived from the radius-luminosity scaling relation, and also yields
$\pm 0.3$ dex or a factor $\sim 2$ uncertainty in this black hole mass estimates,
$M_{\rm BH,FeK\alpha}$, including the measurement errors of the line widths in the data.
As in the case of NGC 7469,
in which the FWHM of the neutral FeK$\alpha$ line is
significantly larger than that of the broad H$\beta $ line,
 $M_{\rm BH,FeK\alpha}$ could be sometimes overestimated.

\subsection{Comparison with Other Black Hole Mass Indicators}
Finally, 
the black hole mass $M_{\rm BH,FeK\alpha}$
estimated from the neutral FeK$\alpha $ line core
is compared with
other estimates of the black hole mass for the same target
to examine the consistency between them.
For the type-1 AGNs,
we compared $M_{\rm BH,FeK\alpha}$ with 
the broad emission-line reverberation mass, $M_{\rm BH,rev}$,
and for the type-2 AGNs, 
we compare it with the mass $M_{\rm BH,H_2O}$ based on the H$_2$O maser
and the single-epoch mass estimate $M_{\rm BH,pol}$
based on the polarized broad Balmer lines.
The $M_{\rm BH,pol}$ is calculated in the same way as the $M_{\rm BH,FeK\alpha}$,
except that the radius-luminosity scaling relation
of the broad H$\beta $ emission line presented by \citet{kosh+14} is used.
The $M_{\rm BH,rev}$, $M_{\rm BH,H_2O}$, $M_{\rm BH,pol}$,
and their references are listed in Tables \ref{data_sy1} and \ref{data_sy2}.

In Figure \ref{mbh_Sy1_Sy2}, $M_{\rm BH,FeK\alpha}$
is plotted against 
other black hole mass indicators such as
$M_{\rm BH,rev}$, $M_{\rm BH,H_2O}$, and $M_{\rm BH,pol}$
for the type-1 and type-2 target AGNs.
We then calculated the Pearson's correlation coeffient,
and also performed linear regression analysis
to $M_{\rm BH,FeK\alpha}$ and the other black hole mass indicators.
The resultant values are listed in Table \ref{fit_results},
and the best-fit linear regression lines are also presented
in Figure \ref{mbh_Sy1_Sy2}.

For the type-1 AGNs,
$M_{\rm BH,FeK\alpha}$ and $M_{\rm BH,rev}$ 
are consistent within $\pm 2\sigma $ except for NGC~7469.
The best-fit linear regression where the slope is fixed to unity
is estimated as
{\footnotesize
\begin{equation}
\log (M_{\rm BH,FeK\alpha}/10^{7.5}M_{\odot}) = -0.15\ (\pm 0.11) + \log (M_{\rm BH,rev}/10^{7.5}M_{\odot})
\label{eqn_feka_rev2}\ ,
\end{equation}}
which indicates that the systematic difference between them is
within a factor of several tens percent.
However,
the correlation between $M_{\rm BH,FeK\alpha}$ and
$M_{\rm BH,rev}$ is not clear as with the preceding study
\citep{2011SCPMA..54.1354J}.
Excluding the clear outlier NGC 7469,
the Pearson's correlation coefficient is calculated
as $R=0.701$, which indicates that the confidence level
of the correlation is less than $90$\%.
The best-fit linear regression for them
is estimated as
{\footnotesize
\begin{equation}
\log (M_{\rm BH,FeK\alpha}/10^{7.5}M_{\odot}) = -0.19\ (\pm 0.17) + 1.52\ (\pm 0.92) \times \log (M_{\rm BH,rev}/10^{7.5}M_{\odot})
\label{eqn_feka_rev}
\end{equation}}
with an additional scatter in both directions
for the reduced $\chi ^{2}$ to achieve unity.
The large uncertainty in the slope of the best-fit linear regression
also indicates no or a weak correlation between them.

For the type-2 AGNs,
$M_{\rm BH,FeK\alpha}$ are also consistent with $M_{\rm BH,pol}$ within $\pm$2$\sigma$.
The best-fit linear regression for $M_{\rm BH,FeK\alpha}$ and $M_{\rm BH,pol}$
is estimated as
{\footnotesize
\begin{equation}
\log (M_{\rm BH,FeK\alpha}/10^{7.5}M_{\odot}) = -0.13\ (\pm 0.13) + 1.05\ (\pm 0.25) \times \log (M_{\rm BH,pol}/10^{7.5}M_{\odot})\ ,
\label{eqn_feka_pol}
\end{equation}}
which indicates that the systematic difference between
$M_{\rm BH,FeK\alpha}$ and $M_{\rm BH,pol}$ is also
within a factor of several tens percent.
In addition,  $M_{\rm BH,FeK\alpha}$ and  $M_{\rm BH,pol}$ are well correlated
and show an approximately proportional relationship, since
the slope of the best-fit linear regression
is positive more than $3\sigma $ uncertainties and close to unity.
The Pearson's correlation coefficients are calculated as $R=0.831$,
which indicates that the confidence level
of the correlation is more than $95$\%.

In contrast to the good agreement with $M_{\rm BH,pol}$,
$M_{\rm BH,FeK\alpha}$ seems to be systematically larger than $M_{\rm BH,H_2O}$
although they show a clear correlation
with the Pearson's correlation coefficient
of $R=0.960$ indicating the confidence level of more than $95$\%.
The best-fit linear regression is estimated as
{\footnotesize
\begin{equation}
\log (M_{\rm BH,FeK\alpha}/10^{6.5}M_{\odot}) = 0.70\ (\pm 0.12) + 1.57\ (\pm 0.33) \times \log (M_{\rm BH,H_2O}/10^{6.5}M_{\odot})\ ,
\label{eqn_feka_h2o}
\end{equation}}
which indicates that 
$M_{\rm BH,FeK\alpha}$ is systematically larger than $M_{\rm BH,H_2O}$
by about a factor of $\sim 5$.
We note that $M_{\rm BH,pol}$ is also larger
than $M_{\rm BH,H_2O}$ by a similar factor
for the three targets, NGC 1068, NGC 4388, and Circinus, 
for which both $M_{\rm BH,H_2O}$ and $M_{\rm BH,pol}$ are estimated.

To summarize the results,
$M_{\rm BH,FeK\alpha}$ is consistent with
$M_{\rm BH,rev}$ for most of the type-1 AGNs
and with $M_{\rm BH,pol}$ for all of the type-2 AGNs.
$M_{\rm BH,FeK\alpha}$ is correlated well
with $M_{\rm BH,pol}$ and $M_{\rm BH,H_2O}$
for the type-2 AGNs,
although $M_{\rm BH,H_2O}$ is a systematic smaller than the others.
These results suggest that
the black hole mass estimate using the neutral FeK$\alpha$ line
is a potential indicator of the black hole mass,
especially for obscured AGNs.

\section{DISCUSSION}

\subsection{Systematic Difference of the H$_2$O maser mass from the Other Black Hole Mass Indicators}
The systematic difference of $M_{\rm BH,H_2O}$
implies an ambiguous issue.
Since $M_{\rm BH,H_2O}$ is considered to be the most
precise estimate for the black hole mass, then,
through the relations of $M_{\rm BH,H_2O}$
with $M_{\rm BH,FeK\alpha}$ and $M_{\rm BH,pol}$,
another reliable estimate $M_{\rm BH,rev}$
is suspected to considerably overestimate the black hole mass
that is difficult to be accounted for
by the uncertainties in the virial factor $f$
($5.5$ in this paper and presented by Onken et al. 2004;
$5.2$ by Woo et al. 2010;
$2.8$ by Graham et al. 2011; $4.31$ by Grier et al. 2013).
In this section, 
the systematic differences between
those black hole mass estimates
and the possible origins of these differences are discussed.

In fact, 
the systematic difference between $M_{\rm BH,pol}$ and $M_{\rm BH,H_2O}$
has been indicated by preceding studies.
\citet{2010ApJ...721...26G} measured
the stellar velocity dispersion $\sigma _{\ast}$ 
in the centers of the galaxies hosting type-2 AGNs
whose $M_{\rm BH,H_2O}$ was obtained,
and found that $M_{\rm BH,H_2O}$ tended to be smaller
than the black hole mass that was estimated from $\sigma _{\ast}$ 
and the $M_{\rm BH}$--$\sigma _{\ast}$ relation of 
local elliptical galaxies.
On the other hand, \citet{2008A&A...488..113Z}
estimated the radius of the BLR from the [\ion{O}{3}] luminosity
via the luminosity scaling relation of the BLR
to estimate $M_{\rm BH,pol}$ of 12 type-2 AGNs,
and found that $M_{\rm BH,pol}$ tended to be larger
than the black hole mass that was estimated from $\sigma _{\ast}$ 
via the $M_{\rm BH}$--$\sigma _{\ast}$ relation.
According to these results, $M_{\rm BH,pol}$ would
be systematically larger than $M_{\rm BH,H_2O}$,
as is the case in this study.

Recently, \citet{2014ApJ...789...17H} presented that
the virial factor $f$ of the reverberation-mapped AGN
is systematically different between two bulge types of the host galaxies,
as $f=6.3\pm 1.5$ for classical bulges and ellipticals,
and $f=3.2\pm 0.7$ for psuedobulges. In fact,
the black hole mass of the AGNs with psuedobulges
tends to be smaller than that with classical bulges and ellipticals
as shown in their Figure 2.
\citet{2010ApJ...721...26G} also argued that
the mass of the black hole hosted by later-type and lower-mass galaxies
tended to be less massive with larger scatter than that estimated from
the $M_{\rm BH}$--$\sigma _{\ast}$ relation for the elliptical galaxies.
Since our type-2 AGN targets with the H$_2$O maser observation
tend to have a less massive black hole than the others
if $M_{\rm BH,H_2O}$ is correct,
then, it is possible that 
the systematic difference of
the target selection
partly contributes to the systematic difference
between $M_{\rm BH,H_2O}$ and the other black hole mass indicators.

Contrary to these studies,
\citet{2011ApJ...727...20K} estimated the radius of the BLR from
the absorption-corrected luminosity in the 2--10 keV X-ray band
$L_{2-10 {\rm keV}}$
via the luminosity scaling relation of the BLR
to estimate $M_{\rm BH,pol}$ of 4 type-2 AGNs
whose $M_{\rm BH,H_2O}$ was obtained by them,
and presented that $M_{\rm BH,H_2O}$ and $M_{\rm BH,pol}$
were consistent with each other.
In fact, the three targets, NGC 1068, NGC 4388, and Circinus,
for which the large systematic difference
between $M_{\rm BH,pol}$ and $M_{\rm BH,H_2O}$ is presented in this paper,
are included in the targets of \citet{2011ApJ...727...20K}.
The reason for the discrepancy between these studies is uncertain,
but a difficulty in estimating the luminosity of
the central engine $L_{2-10 {\rm keV}}$ for most of the targets
with very large X-ray absorbing column density
could be suspected,
although it was estimated with particular attention.

Assuming both $M_{\rm BH,rev}$ and $M_{\rm BH,H_2O}$
are reliable estimates for the black hole mass,
$M_{\rm BH,FeK\alpha}$ and $M_{\rm BH,pol}$
are considered to overestimate the black hole mass
in type-2 AGNs.
The systematically larger $M_{\rm BH,FeK\alpha}$
for the type-2 AGNs implies that
the width of the neutral FeK$\alpha $ line core
is systematically larger for the large inclination angle for type-2 AGNs.
If the emitting region of the neutral FeK$\alpha $ line core
shows an equatorial motion,
then the virial factor $f$ in Equation (\ref{eqn_virial}) decreases
as the inclination angle increases.
For example, $f=2$ at the edge-on view,
which is smaller than $f=5.5$ for type-1 AGNs,
is derived assuming an equatorial circular motion.
However, it is not so small as to account for
the difference of about a factor of $\sim 5$.
Outflow motions toward the equatorial plane
might broaden the width of
the neutral FeK$\alpha $ line core for type-2 AGNs.

On the other hand,
the systematically larger $M_{\rm BH,pol}$ for the type-2 AGNs
cannot be attributed to the inclination of AGNs
from the point of view of the unified model of AGNs;
when the broad emission line is scattered to the observer
at the polar region well above the height of the dust torus,
the viewing angle of the BLR from the scattering region
would be similar to that from the observer for type-1 AGNs.
Non-virial motions such as outflows in the scattering region
are possible to be contributed, as \citet{2007Natur.450...74Y}
proposed to explain the broad H$\alpha $ emission line
in the polarized spectrum of the quasar PG 1700$+$518.
However, a common or closely related mechanism to broaden
the widths of both of the neutral FeK$\alpha $ line core and
the polarized broad Balmer emission lines
systematically for type-2 AGNs is required
to explain the approximate agreement between
$M_{\rm BH,FeK\alpha}$ and $M_{\rm BH,pol}$.

\subsection{Location of the Neutral FeK$\alpha $ Line Core Emitting Region}
As described in Section 3,
the FWHM of the neutral FeK$\alpha $ line core falls
between the FWHM of the broad Balmer emission lines and
the corresponding values at the dust reverberation radius
in most cases,
which indicates that the emitting region of
its major fraction is located
between the outer BLR and the inner dust torus.
It is suggested by previous studies for some of the target AGNs.
For NGC 4151, \citet{2002PASJ...54..373T} compared
the variation of the continuum and the excess flux
around the neutral FeK$\alpha $ line and presented that
the emitting region of the excess flux has an extent of $10^{17}$ cm,
or 40 light days, which is consistent with or slightly smaller than
the dust reverberation radius \citep{kosh+14}.
For NGC 5548, \citet{2010ApJ...710.1228L} reported that
the emitting region of the neutral FeK$\alpha $ line
was located between the two reverberation radii
of the broad H$\beta $ emission line and the near-infrared dust emission
based on its flux variation and line width.

The BLR and the dust torus are considered not to be decoupled
from each other but to constitute a continuous structure,
and the transition zone of the BLR and the dust torus
would be located between the two reverberation radii
\citep{2008ApJ...685..160N,2009ApJ...700L.109K,kosh+14}.
In addition, recent reverberation observations 
of the optical \ion{Fe}{2} emission lines
presented that its reverberation radius
was comparable to or at most twice
that of the broad H$\beta $ emission line
\citep{2013ApJ...769..128B,2014ApJ...783L..34C},
which shows the presence of low ionization iron atoms there.

\begin{figure}
\epsscale{0.7}
\plotone{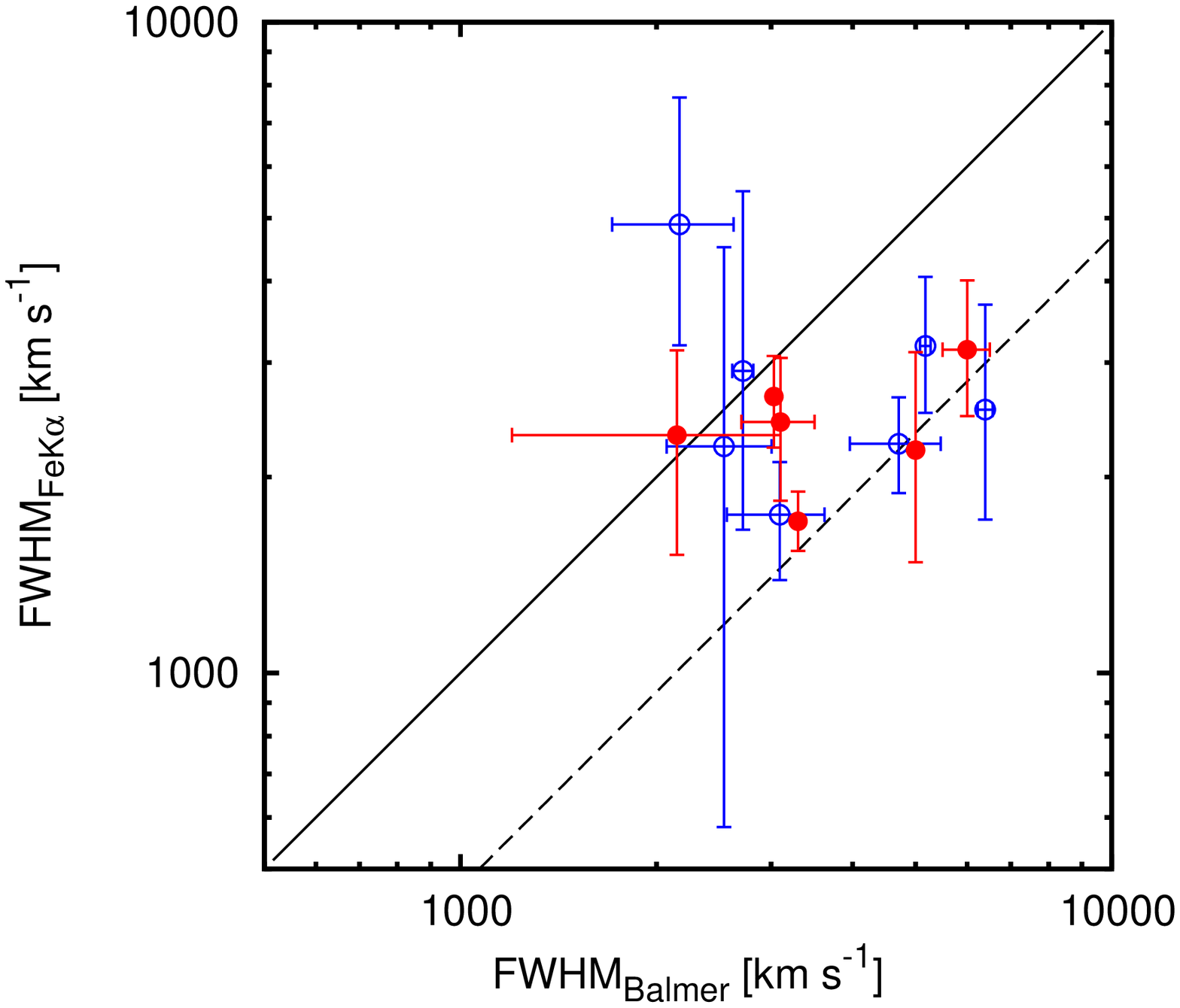}
\caption[The FWHM]{
The FWHM of the neutral FeK$\alpha $ line core
against that of the broad Balmer emission lines.
The open circles (colored blue in the online version)
represent the broad H$\beta $ emission line for the type-1 AGNs,
and the filled circles (colored red in the online version)
represent the polarized broad Balmer emission lines for the type-2 AGNs.
The solid line represents unity,
and the dashed line represents 
the FWHM at the $K$-band dust reverberation radius
\citep[][]{kosh+14}.
(A color version of this figure is available in the online journal.)
}\label{fwhm_fwhm}
\end{figure}

\begin{figure}
\epsscale{0.7}
\plotone{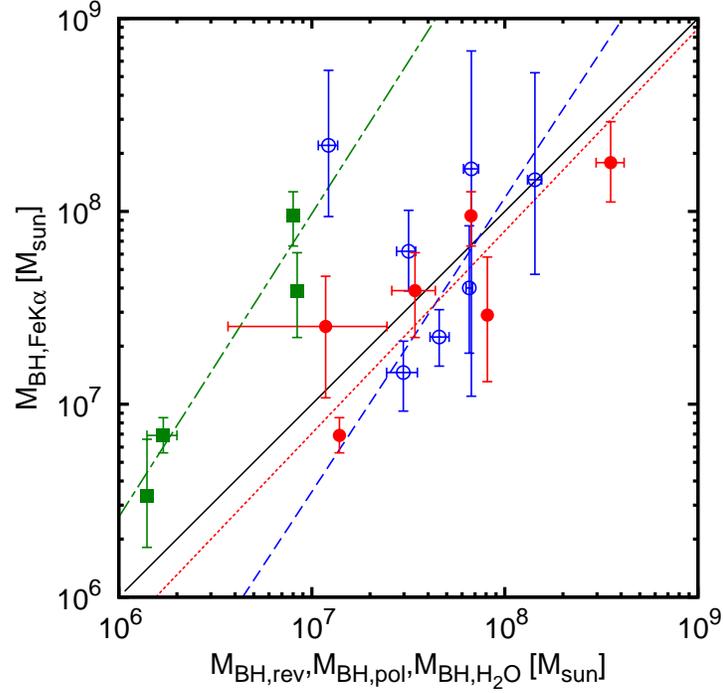}
\caption[The black hole mass estimates of the Seyfert galaxies]{
The black hole mass estimate based on the neutral FeK$\alpha $ line core,
$M_{\rm BH,FeK\alpha}$, and other black hole mass estimates.
The open circles (colored blue in the online version)
represent the broad emission-line reverberation mass, $M_{\rm BH,rev}$
for the type-1 AGNs.
The filled boxes (colored green in the online verseion)
represent the black hole mass based on the VLBI observation
of the H$_2$O maser, $M_{\rm BH,H_2O}$,
and the filled circles (colored red in the online version)
represent the single-epoch mass estimate based on the polarized broad
Balmer emission lines, $M_{\rm BH,pol}$, for the type-2 AGNs.
The solid line represents unity,
and the dashed , dot-dashed, and dotted lines
(colored blue, green, and red, respectively, in the online version)
represent the best-fit linear regression lines of $M_{\rm BH,FeK\alpha}$
to $M_{\rm BH,rev}$, $M_{\rm BH,H_2O}$, and $M_{\rm BH,pol}$, respectively.
(A color version of this figure is available in the online journal.)
}\label{mbh_Sy1_Sy2}
\end{figure}

If the FeK$\alpha $ emitting region constitutes
a tight stratified structure with the BLR and the dust torus,
a good correlation between the FWHMs of
the neutral FeK$\alpha $ line and the broad Balmer emission lines
is expected.
However, as shown in Figure \ref{fwhm_fwhm},
there appears no clear correlation between them
as has been reported by previous studies
\citep[][]{2006MNRAS.368L..62N,2010ApJS..187..581S,2011ApJ...738..147S}.
A simple explanation for it would be that the neutral FeK$\alpha$ line
is produced in different origins such as the outer accretion disk,
the BLR, and the innermost dust torus,
and the mixing ratio is different from target to target
\citep[e.g.,][]{2004ApJ...604...63Y,2006MNRAS.368L..62N,2008MNRAS.389L..52B,2010ApJ...710.1228L,2013A&A...549A..72P}.
On the other hand,
\citet[][]{2011ApJ...738..147S}
suggested that the neutral FeK$\alpha$ line may originate
from a universal region at the same radius with respect
to the gravitational radius of the central black hole,
because its FWHM is almost constant and independent
of the black hole mass.
In any case, those uncertainties
will result in the possible
error of the black hole mass estimated from
the neutral FeK$\alpha$ line.

In this study, We used the best spectral resolution data
currently available,
but the spectral resolution and sensitivity are still limited.
As a result, the number of the targets is small
and the relatively large uncertainties remain in the FWHM data,
which would make it difficult to examine
the location of the neutral FeK$\alpha$ line emitting region
in more detail.
In the near future,
the substantial progress is expected
by the {\it ASTRO-H} X-ray satellite \citep{2010SPIE.7732E..27T},
which is capable of unprecedented energy-resolution spectroscopy
with superior sensitivity at FeK$\alpha $ band.
It will enable us to study more detail on the origin of the neutral FeK$\alpha $ line core
by examining the line profile of the neutral FeK$\alpha $ line
and its time variation,
then it will become able to measure the black hole mass
with improved accuracy
for a large number of obscured AGNs.
The hard X-ray luminosity with precise absorption correction
is obviously valuable
for the isotropic luminosity indicator
because the scaling relations of the reverberation radii
to the hard X-ray luminosity are tighter than to the [\ion{O}{4}] luminosity
\citep{kosh+14}.

\acknowledgments

We thank T. Kawaguchi, M. Yoshida, K. Kawabata, and Y. Fukazawa
for useful discussions and comments.
We also thank the anonymous referee
for valuable comments to improve the manuscript.
This research has been partly supported by
the Grants-in-Aid of Scientific Research (22540247 and 25287062)
of the Ministry of Education, Science, Culture and Sports of Japan,
and has made use of the NASA/IPAC Extragalactic Database (NED),
which is operated by the Jet Propulsion Laboratory,
California Institute of Technology,
under contract with the National Aeronautics and Space Administration. 

\hspace{10cm}

\clearpage

\clearpage

\clearpage

\begin{deluxetable}{lcccccc}
\tablewidth{0pt}
\tabletypesize{\footnotesize}
\tablecaption{List of the target type-1 AGNs}
\tablehead{
\colhead{Object}     & 
\colhead{FWHM$_{\rm FeK\alpha }$} &
\colhead{$\log \ L_{\rm [OIV]}$} & 
\colhead{$M_{\rm BH,FeK\alpha}$} &
\colhead{FWHM$_{\rm H\beta }$} & \colhead{$M_{\rm BH,rev}$} & \colhead{refs\tablenotemark{a}} \\
                     & 
\colhead{km s$^{-1}$}         &
\colhead{erg s$^{-1}$}       &
\colhead{$10^{7}M_{\odot }$} &
\colhead{km s$^{-1}$}      & \colhead{$10^{7}M_{\odot }$}& }
\startdata
3C 120         & $2230^{+2280}_{-1650}$ & 42.42   & $17^{+51}_{-16}$  & 
                 $2539\pm 466$                      & $6.70^{+0.60}_{-0.60}$ & 1 \\
NGC 3516       & $3180^{ +880}_{ -670}$ & 40.95   & $6.2^{+3.9}_{-2.3}$ &
                 $5175\pm  96$                      & $3.17^{+0.28}_{-0.42}$ & 2 \\
NGC 3783       & $1750^{ +360}_{ -360}$ & 40.73   & $1.46^{+0.66}_{-0.54}$ &
                 $3093\pm 529$                      & $2.98^{+0.54}_{-0.54}$ & 3 \\
NGC 4151       & $2250^{ +400}_{ -360}$ & 40.66   & $2.23^{+0.86}_{-0.66}$ &
                 $4711\pm 750$                      & $4.57^{+0.57}_{-0.47}$ & 4 \\
NGC 5548       & $2540^{+1140}_{ -820}$ & 40.96   & $4.0^{+4.4}_{-2.2}$ &
                 $6396\pm 167$                      & $6.54^{+0.26}_{-0.25}$ & 5 \\
Mrk 509        & $2910^{+2590}_{-1250}$ & 41.85   & $15^{+38}_{-10}$ &
                 $2715\pm 101$                      & $14.3^{+1.2}_{-1.2}$   & 3 \\
NGC 7469       & $4890^{+2770}_{-1700}$ & 41.30   & $22^{+32}_{-13}$ &
                 $2169\pm 459$                      & $1.22^{+0.14}_{-0.14}$ & 3   
\enddata
\label{data_sy1}
\tablerefs{(1) \citet{grie+12}; (2) \citet{denn+10}; (3) \citet{2004ApJ...613..682P}; (4) \citet{bent+06}; (5) \citet{bent+07}.
}
\tablenotetext{a}{References for FWHM of the broad H$\beta $ emission line, FWHM$_{\rm H\beta}$,
and the black hole mass from the broad emission-line reverberation, $M_{\rm BH,rev}$.}
\end{deluxetable}

\begin{deluxetable}{lcccccccc}
\tablewidth{0pt}
\tabletypesize{\footnotesize}
\tablecaption{List of the target type-2 AGNs}
\tablehead{
\colhead{Object}     & 
\colhead{FWHM$_{\rm FeK\alpha }$} &
\colhead{$\log \ L_{\rm [OIV]}$\tablenotemark{a}} & \colhead{$M_{\rm BH,FeK\alpha}$} &
\colhead{FWHM$_{\rm pol}$} & \colhead{refs\tablenotemark{b}} & \colhead{$M_{\rm BH,pol}$} &
\colhead{$M_{\rm BH,H_{2}O}$} & \colhead{refs\tablenotemark{c}} \\
                     & 
\colhead{km s$^{-1}$} & \colhead{erg s$^{-1}$} & \colhead{$10^{7}M_{\odot }$} &
\colhead{km s$^{-1}$} & & \colhead{$10^{7}M_{\odot }$} & \colhead{$10^{7}M_{\odot }$} & }
\startdata
NGC 1068       & $2660^{ +410}_{ -440}$ & 41.67 & $9.5^{+3.2}_{-2.9}$ &
                 $3030$                     & 1 & $6.7$                  &
                                                  $0.80^{+0.03}_{-0.03}$ & 8 \\
NGC 2110       & $2320^{ +810}_{ -800}$ & 40.76 & $2.5^{+2.1}_{-1.5}$ &
             $2150\pm 950$\tablenotemark{d} & 2 & $1.2^{+1.3}_{-0.8}$ &
                                                  \nodata                & \nodata  \\
Mrk 3          & $3140^{ +870}_{ -660}$ & 41.93 & $18^{+11}_{-7}$        &
                 $6000\pm 500$              & 3 & $35^{+6}_{-6}$         &
                                                  \nodata                & \nodata \\
NGC 4388       & $2430^{ +620}_{ -590}$ & 41.05 & $3.9^{+2.2}_{-1.7}$    &
                 $3100\pm 400$              & 4 & $3.4^{+0.9}_{-0.8}$ &
                                                  $0.84^{+0.02}_{-0.02}$ & 9       \\
NGC 4507       & $2200^{ +910}_{ -720}$ & 40.97 & $2.9^{+2.9}_{-1.6}$    &
                 $5000$                  & 5, 6 & $8.1$                  &
                                                  \nodata                & \nodata \\
NGC 4945       & $2780^{+1110}_{ -740}$ & 38.69 & $0.34^{+0.32}_{-0.16}$ &
                 \nodata              & \nodata & \nodata        &
                                                  $0.14$                 & 10 \\
Circinus       & $1710^{ +190}_{ -170}$ & 40.16 & $0.69^{+0.16}_{-0.13}$ &
                 $3300$                     & 7 & $1.4$                     &
                                                  $0.17^{+0.03}_{-0.03}$ & 11   
\enddata
\label{data_sy2}
\tablerefs{(1)\citet{1998ApJ...499..134N}; (2)\citet{2007ApJ...668L..31M}; (3) \citet{1995ApJ...440..565T}; (4) \citet{1996MNRAS.281.1206Y}; (5) \citet{2000ApJ...540L..73M}; (6) \citet{2011ApJ...738..147S}; (7) \citet{1998A&A...329L..21O}; (8) \citet{2003A&A...398..517L}; (9) \citet{2011ApJ...727...20K}; (10) \citet{1997ApJ...481L..23G}; (11) \citet{2003ApJ...590..162G}.
}
\tablenotetext{a}{The same distance as that of the $M_{\rm BH,H_{2}O}$ reference is adopted when available.}
\tablenotetext{b}{References for FWHM of the polarized broad Balmer emission lines, FWHM$_{\rm pol}$.}
\tablenotetext{c}{References for the black hole mass based on the H$_2$O maser observation,
$M_{\rm BH,H_{2}O}$.}
\tablenotetext{d}{The average FWHM of the polarized H$\beta $ emission line
and the core component of the polarized H$\alpha $ emission line.}.
\end{deluxetable}

\begin{deluxetable}{lcccccccc}
\tablewidth{0pt}
\tablecaption{Correlation coefficients and linear regressions to $M_{\rm BH,FeK\alpha }$}
\tablehead{
\colhead{$M_{\rm BH,ref}$} & $n$\tablenotemark{a} & $R$\tablenotemark{b} & \colhead{$\log M_{\rm BH,0}$}\tablenotemark{c} & $a$\tablenotemark{c} & $b$\tablenotemark{c} \\
 &  & & \colhead{[$M_{\odot}$]} &  & 
}
\startdata
$M_{\rm BH,rev}$\tablenotemark{d}    & $6$ & $0.701$ & $7.5$ & $-0.19 \pm 0.17$ & $1.52 \pm 0.92$ \\
                    &     &         &                        & $-0.15 \pm 0.11$ & $1.0$ (fixed)   \\
$M_{\rm BH,pol}$                     & $6$ & $0.831$ & $7.5$ & $-0.13 \pm 0.13$ & $1.05 \pm 0.25$ \\
                    &     &         &                        & $-0.12 \pm 0.11$ & $1.0$ (fixed)   \\
$M_{\rm BH,H_{2}O}$                  & $4$ & $0.960$ & $6.5$ & \ \, $0.70 \pm 0.12$ & $1.57 \pm 0.33$ & \\
                    &     &         &                        & \ \, $0.73 \pm 0.14$ & $1.0$ (fixed)   
\enddata
\label{fit_results}
\tablenotetext{a}{The number of the data.}
\tablenotetext{b}{The Pearson's correlation coefficient.}
\tablenotetext{c}{The fitted model is $\log (M_{\rm BH,FeK\alpha }/M_{\rm BH,0})=a + b\times \log (M_{\rm BH,ref}/M_{\rm BH,0})$, where $a$ and $b$ are the parameters to be fitted.}
\tablenotetext{d}{NGC 7469 was excluded from the calculations of the correlation coefficient and the linear regression.}
\end{deluxetable}

\end{document}